\documentstyle[11pt]{article}
\begin{document}
\begin{center}
\large{
Rational Galaxy Structure and its Disturbance
}
\normalsize\\
Jin He \\
Department of Physics, Huazhong University of Science and Technology, \\Wuhan, Hubei 430074, China\\
E-mail:mathnob@yahoo.com\\
\mbox{   }
\end{center}
\large {{\bf Abstract} } \normalsize
Why is there little dust in elliptical galaxies?
Firstly, the universe is rational and the rationality is the proportional density distribution in any independent galaxy. Proportion means that the density distribution of stars is orderly. For example, there are four giants standing in array. Their heights are A, B, C, D, respectively, and A, B stand in the first row from left to right, C, D in the second row from left to right. Proportion requires that A divided by C be equal to B divided by D. Rational structures are smooth 2- (or 3-) dimensional structures, which are either circularly
(spherically) symmetric with respect to the center point, or bilaterally symmetric.
Secondly, spiral galaxy arms are linearly shaped, which are neither circularly symmetric nor bilaterally symmetric. Therefore, arms are not rational structures. Arm patterns exist only in spiral galaxies and they are weak compared with the main disk structure. Therefore, the presence of arm structure is the disturbance to the rational structure. Because arm patterns exist only in spiral galaxies and only spiral galaxies present dust, I come to the critical answer to the above question: any disturbance to rational structure produces cosmic dust.
From now on we call the disturbance to rational structure the cosmic impulse. Therefore, the universe is originated not only rationally but also impulsively.
\\
\\
keywords: Methods : Analytical -- Galaxies : General
\\
\\
\section{ The Origin of Cosmic Dust  }

\subsection{ The Power which Governs Everything: Gravitation }

Scientists have fully proved that there exist only four forces among
particles: electromagnetic, weak nuclear, strong nuclear, and
gravitational. The nuclear forces are short-ranged while the
electromagnetic force is long-ranged. Each of the three has two
contradictory aspects of attracting and repelling and, generally, has
no net effect in the macroscopic world due to offsetting effect.
Gravitational force, however, has no contradictory aspects. Gravity
has the only effect of attraction and, therefore, can not offset
itself. Because of this, the true origin of natural structure is the
gravitational force.

The origin of natural structure can not be other forces. Modern
science has fully proved that independent system of microscopic
particles combined by electromagnetic force or nuclear forces
inevitably moves towards chaotic state rather than orderly one. This
is the principle of entropy increase, which is well known for
scientists. Therefore, if there were no gravitational force then the
whole universe would be simply uniform gas without structure.
However, there exist in the macroscopic world such orderly
structures large as galaxies and small as stars, planets, plants,
animals, and even human beings. Therefore, varied kinds of
macro-world structures result from the struggling of the
gravitational force against the electromagnetic and nuclear forces.

Unfortunately, gravity is very and very weak. For example, it is
$0.0000 \cdots 00001 $ (where 40 zeros are after the decimal point)
times weaker than the electricity. Only the Earth, Moon, Sun and so
on present gravity. There is no slightest gravity between cars or
human bodies. Therefore, human beings suffer insurmountable
difficulty to experimentally study gravity. Because human beings are
insignificant, it is impossible to do physical experiments on
such macroscopic systems as the solar system or galaxies.
However, we can use man-made telescopes to take images of
large-scale material systems such as galaxies. We can analyze the
images.

\subsection{ The Origin of Galaxies: Rational Gravity }

The usually familiar gravity refers to the force which exerts
between Earth and Moon or between Sun and Earth. These examples of
gravity are the interaction between two bodies. As for the behavior
of gravity exerted on many bodies, the solar system can not be the
example for us to study such behavior. However, each galaxy is
composed of billions of stars, which demonstrates the gravitational
interaction among many bodies. Galaxy images show that each galaxy
has a center. Star density at the center is the highest. From the
center outward, the density is smaller and smaller and presents a
regular pattern, known as galaxy structure. The principle behind the
formation of galaxy structures is the demonstration of gravitational
interaction in many-body systems.

Then, what is the behavior of gravitational interaction in many-body
systems? we pioneered the study on galaxy structures (He (2003, 2004, and 2006), He \& Yang (2006)) and
the study shows that stars in any galaxy are controlled by a very
simple orderly force involving many-bodies: proportion. Because
solar system is just a point at the Milky Way galaxy, the proportion
force reduces to Newtonian gravity between two bodies.

Proportion means that the distribution of matters in the universe is
orderly.
For example, there are four giants standing in array. Their heights are A, B, C, D, respectively, and A, B stand in the first row from left to right, C, D in the second row from left to right. According to the common view, the four giants
can have any heights and can stand at any position. That is why
current scientific theories are incompetent and can not
explain the origin of natural structures. They can not provide any
basic principle to resolve the motion of the most simple gravitational systems
(such as interactional free three-bodies). However, the universe is
orderly. The orderly force at the largest scales requires that the
distribution of heights is in proportion. In other words, A divided
by B is equal to C divided by D. This means that A divided by C is
equal to B divided by D. If there are nine giants standing in array,
then the ratios of heights from neighboring two rows are constant
(proportion rows). Similarly, the ratios of heights from neighboring
two lines are constant (proportion lines). In this way are galaxies
created.

The above-said rows and lines are all straight (proportion lines).
But each galaxy is a regional distribution of matters in the
universe. Therefore, the proportion lines of each galaxy are curved
but the rows and columns still cross at each other vertically and
they form the net of orthogonal curves.

In general, a distribution of similar bodies is called the rational
structure if its density varies proportionally along some particular
net of orthogonal curves. Therefore, independent
galaxies are all rational structures. The force which leads to the
rational structure is called the rational force, i.e., the
proportion force which is the demonstration of gravity at
large-scale and many-body system. That is, the universe is rational.
This is an important discovery in science.

\section{ Two Examples of Rational Structure  }

\subsection{ Logarithmic Density of Galaxy Structure }

Now we start the scientific and mathematical investigation into
galaxy structures. A galaxy is a distribution of stars. But we can
not see individual stars on a galaxy image.  A galaxy image is the
distribution of star densities. Therefore, we use a mathematical
function to describe a distribution of densities. Because spiral
galaxies are planar, we use a function of two variables, $x,y$, to
describe the stellar distribution of a spiral galaxy:
\begin{equation}
 \rho (x,y)
\end{equation}
where  $x,y$ is the rectangular Cartesian coordinates on the spiral
galaxy plane. The coordinate origin is the galaxy center. Therefore,
$ \rho (0,0)$ is the stellar density at the galaxy center. We want
to study the ratio of the density $ \rho _2$  to the density $ \rho
_1$ at two positions 2 and  1 respectively:
\begin{equation}
 \rho _2 / \rho _1.
\end{equation}
In fact, the logarithm of the ratio divided by the distance $s$
between the two positions is approximately the directional
derivative of the logarithmic density ($ f(x,y) = \ln \rho (x,y)$)
along the direction of the two positions:
\begin{equation}
(\ln(\rho _2/\rho _1))/s= (\ln\rho _2-\ln\rho _1)/s \approx \frac
{\partial f}{\partial s}
\end{equation}
There is no systematic mathematical theory on ratios. Therefore, we
from now on, study the logarithmic function $f(x,y)$ instead of the
density function $ \rho (x,y)$:
\begin{equation}
f(x,y)=\ln \rho(x,y).
\end{equation}

\subsection{ Description of a Net of Orthogonal Curves }

The following equation
\begin{equation}
 \left\{
 \begin{array}{l}
x=x(\lambda,\mu) \\
y=y(\lambda, \mu)
\end{array}
\right.
\end{equation}
tells us how to describe a net of curves by employing mathematics.
Given two functions, $x(\lambda,\mu)$, $y(\lambda,\mu)$, you have
the transformation between the curvilinear coordinates
$(\lambda,\mu)$ and  the rectangular Cartesian coordinates $(x, y)$,
i.e., the formula (5). It describe a net of curves. Letting the
second parameter $\mu $ be a constant, you have a curve (called a
row curve, i.e., the proportion row defined in the following). That
is, the formula (5) is a curve with its parameter being $\lambda
$. For the different values of the constant $\mu $, you have a set
of ``parallel'' rows. On the other hand, you let the first parameter
$\lambda $ be a constant then you have another curve (called a
column curve). That is, the formula (5) is a curve with its
parameter being $\mu $. For the different values of the constant
$\lambda $, you have a set of ``parallel'' columns.

However, The row curves and the column curves are not necessarily
orthogonal to each other. The following equation is the necessary
and sufficient condition for the net of curves to be orthogonal:
\begin{equation}
\frac {\partial x}{\partial \lambda }\frac {\partial x}{\partial \mu
}+ \frac {\partial y}{\partial \lambda }\frac {\partial y}{\partial
\mu} \equiv 0.
\end{equation}

To study the rational structures described in the following, we need
more knowledge of the description of row and column curves. The arc
length of the row curve is $s$ whose differential is the following:
\begin{equation}
ds=  \sqrt {x_\lambda ^{\prime 2} + y_\lambda ^{\prime 2} } d\lambda
=P d\lambda .
\end{equation}
where $P$ is the arc derivative of the row curve:
\begin{equation}
P=s_\lambda ^\prime = \sqrt {x_\lambda ^{\prime 2} +  y_\lambda
^{\prime 2} } .
\end{equation}
The arc length of the column curve is  $t$ whose differential is the
following:
\begin{equation}
dt= \sqrt {x_\mu ^{\prime 2} + y_\mu^{\prime 2} } d\mu =Q d\mu .
\end{equation}
where $Q$ is the arc derivative of the column curve:
\begin{equation}
Q=t_\mu ^\prime =\sqrt {x_\mu ^{\prime 2} +  y_\mu ^{\prime 2}} .
\end{equation}

\subsection{ The Condition of Rational Structure }

The formulas (5) and (6) are the general description of a net of
orthogonal curves, and the formula (1) is the general description
of a distribution of densities (a structure). This paper talks about
rational structure. A distribution of densities is called the
rational structure if its density varies proportionally along some
particular net of orthogonal curves. That is, you walk along a curve
from the net and the ratio of the density on your left side to the
immediate density on your right side is constant along the curve.
However, the constant ratio of this curve is generally different
from the constant ratios of the other curves.

We have shown that a logarithmic ratio of  two densities divided by
the distance between the two positions is approximately the
directional derivative of the logarithmic density along the
direction of the distance. Therefore, we always study the
logarithmic function $f(x,y)$ (see the formula (4)). If we know
the two partial derivatives $\partial f/\partial x $ and $\partial
f/\partial y $ then the structure $f(x,y)$ is found. The partial
derivatives of $f(x,y)$ are the directional derivatives along the
straight directions of the rectangular Cartesian coordinate lines.
However, we are interested in the net of orthogonal curves and what
we look for is the directional derivatives along the tangent
directions of the curvilinear rows and columns. These are denoted
$u(\lambda, \mu)$ and $v(\lambda, \mu)$ respectively.

The condition of rational structure is that $u$ depends only on
$\lambda $ while $v$ depends only on $\mu $:
\begin{equation}
\begin{array}{l}
u=u(\lambda ), \\
v=v(\mu ).
\end{array}
\end{equation}
What a simple condition for the solution of rational structure!

Now we prove the condition. Assume you walk along a row curve. The
logarithmic ratio of the density on your left side to the immediate
density on your right side is approximately the directional
derivative of $f(x,y)$ along the column direction. That is, the
logarithmic ratio is approximately the directional derivative
$v(\lambda, \mu )$. Because $v(\lambda, \mu )$ is constant along the
row curve (rational), $v(\lambda, \mu )$ is independent of $\lambda
$: $v=v(\mu ) $. Similarly, we can prove that $u(\lambda, \mu
)=u(\lambda )$.

\subsection{ The Equation of Rational Structure }

It is known that, given an arbitrary function, we can have its two
partial derivatives. However, given two functions, we may not find
the third function whose partial derivatives are the given
functions. For the given two functions to be some derivatives, a
condition must be satisfied. The condition is the Green's theorem.
In the case of derivatives along orthogonal curves, the Green's
theorem is the following:
\begin{equation}
\frac {\partial }{\partial \mu }(u(\lambda ,\mu) P)- \frac {\partial
}{\partial \lambda }(v(\lambda ,\mu) Q)=0
\end{equation}

In the case of rational structure, directional derivatives are the
functions of the single variables, $\lambda $ and $\mu $
respectively (see (11)). Therefore, the Green's theorem turns out
to be the following which is called the rational structure equation:
\begin{equation}
  u(\lambda ) P_\mu ^\prime -v(\mu )Q_\lambda ^\prime =0
\end{equation}

This equation determines rational structure. To find a rational
structure, generally we are first of all given a net of orthogonal
curves. Accordingly the arc derivatives of both the rows and
columns, $P(\lambda,\mu )$ and $Q(\lambda,\mu )$, are known.
Therefore, the remaining functions , $u(\lambda )$ and $v(\mu )$,
are the only unknowns. Note that the unknowns are the functions of
the single variables, $\lambda $ and $\mu $ respectively. Because
the rational structure equation involves no derivative of the
unknowns, the equation is not a differential equation at all. It is
an algebraic equation and what we need to do is to add two factors
(the two unknowns) to the derivatives of $P$ and $Q$ so that the
rational structure equation holds: factorization!

How simple is the universe constructed!

\subsection{ The First Example: Spiral Galaxy Disk Structure }

Images of spiral galaxies taken with infrared light show that each
spiral galaxy is mainly a disk (the circularly symmetric disk with
respect to the galaxy center, i.e., the disk center) and the disk
light density decreases exponentially outwards along the radial
direction from the center. There are other minor or weak structures
in spiral galaxies. Spiral galaxies gain their name by the fact that
they present more or less spiral structures, known as arms.

The first example of rational structure is the spiral galaxy disk
and is determined by the following net of equiangular spirals (or
called logarithmic spirals) (see Fig. 1):
\begin{equation}
\left\{
\begin{array}{l}
x = e^{d_1\lambda +d_2\mu }\cos(d_3\lambda +d_4\mu ) ,   \\
y = e^{d_1\lambda +d_2\mu }\sin(d_3\lambda +d_4\mu )
\end{array}
\right.
\end{equation}
where $d_1 (>0),d_2 (>0),d_3(<0),d_4(>0)$ are real constants and we
choose
\begin{equation}
 d_3=-d_1d_2/d_4
\end{equation}
so that the curves are orthogonal. The polar angle and polar
distance of the point $(x,y)$ are easily seen,
\begin{equation}
\begin{array}{l}
r = e^{d_1\lambda +d_2\mu } ,\\
\theta =d_3\lambda +d_4\mu .
\end{array}
\end{equation}
The coordinate lines are spiral-shaped and shown in Fig.,1.

The corresponding arc-length derivatives $P,Q$ are
\begin{equation}
\begin{array}{l}
P(\lambda,\mu) = s_\lambda ^\prime =  (d_1/d_4) \sqrt{d_2^2+d_4^2} e^{d_1\lambda +d_2\mu },\\
Q(\lambda,\mu) = t_\mu ^\prime = \sqrt{d_2^2+d_4^2} e^{d_1\lambda +d_2\mu }.\\
\end{array}
\end{equation}
The rational structure equation (13) does help factor out the
required directional derivatives for our spiral galaxy disks,
\begin{equation}
\begin{array}{l}
u(\lambda ) = d_5d_4,\\
v(\mu) =d_5d_2
\end{array}
\end{equation}
where $d_5$ is another constant.

So far, we did not specify the variance domain $S$ on $(\lambda ,
\mu )$ coordinate plane on which the coordinate system (14) is
defined. There are many such domains which are proved in the
Appendix of the paper. Here we simply present one kind of such
domain. We define a constant,
\begin{equation}
\Delta _\lambda =\frac {2\pi d _2}{d _1d _4-d _3d _2} \;(>0),  \\
\end{equation}
The domain is the following:
\begin{equation}
S_{\lambda _1} :  \lambda _1 < \lambda < \lambda _2 (=\lambda _1
+\Delta _\lambda ), \; -\infty < \mu < +\infty .
\end{equation}
where $\lambda _1 $  is arbitrary constant and the length of the
interval $(\lambda _1 , \lambda _2 )$ is
\begin{equation}
 \lambda _2 = \lambda _1 +\Delta _\lambda .
\end{equation}
In the Appendix, we also prove that the spiral curves are
equiangular spirals.

The spiral disk density along all orthogonal coordinate lines can be
found by performing path integrations of the following formulas
\begin{equation}
\begin{array}{l}
df=u ds = u(\lambda) P d \lambda ,\\
df=v dt = v(\mu ) Q d \mu .
\end{array}
\end{equation}
along the row curves, $\mu =$ constant, and the column curves,
$\lambda =$ constant, respectively. Without  loss of generality, we
assume the coordinates $\lambda $ and $\mu $ are defined on the
domain (20). Then starting at the galaxy center, we perform the
path integration of $df = v(\mu ) Qd\mu $ over $[-\infty , \mu ]$ by
taking $\lambda $ to be the constant $\lambda _1 $ to get $f(\lambda
_1 , \mu )$. Then we perform the path integration of $df = u(\lambda
) Pd\lambda $ over $[\lambda _1 , \lambda ]$ by taking $\mu $ to be
an arbitrary constant  to get $f(\lambda  , \mu )$. Finally, we have
the logarithmic function $f(x,y)$ implied by the spiral-shaped
coordinate system. The density distribution $\rho (x,y)$ represents
spiral galaxy disks (we choose $d_5<0$, because light density $\rho
\rightarrow 0$ when $r \rightarrow +\infty $),
\begin{equation}
\begin{array}{l}
f_d =d_5\sqrt{d_2^2+d_4^2}e^{d_1\lambda +d_2\mu },\\
\rho _d =d_0\exp (d_5\sqrt{d_2^2+d_4^2}e^{d_1\lambda +d_2\mu })
\end{array}
\end{equation}
where $d_0$ is the light density (star density) at the galaxy
center. Note that we use the letter $d$ as well as the subscript $d$
for disk parameters and formulas. Similar notations are used for bar
parameters and formulas. Because $ f_d =d_5Q(\lambda ,\mu )$ and
$Q(\lambda ,\mu )$ is uniquely defined over the whole galaxy disk
plane ($(x,y)$ plane, see the Appendix A), $f_d(x,y)$ is uniquely
defined on the same plane. We can see the disk light pattern by
displaying $\rho _d (x,y)$. Because the polar distance is $r = \exp
(d_1\lambda +d_2\mu ) $, the galaxy disk light distribution is
circularly symmetric,
\begin{equation}
\rho _d =d _0 e^{f_d} =d _0 e^{(d _5\sqrt{d _2^2+d _4^2})r}.
\end{equation}
Since galaxy light density $\propto  \rho  $, we have recovered the
known exponential law of spiral galaxy disk (the exponential disk).

The circularly symmetric exponential disk is completely determined
by the value of $d_0$ and the value of $d _5\sqrt{d _2^2+d _4^2} $.
Therefore, given a spiral galaxy disk, that is, given the two
specific values, we can find an infinite number of coordinates
(14) defined on the disk plane which give the same disk structure.
From now on, we choose $\sqrt{d _2^2+d _4^2}=1$ so that the disk
model involves only two variable disk parameters $d_0 (>0)$ and $
d_5 (<0)$,
\begin{equation}
\rho _d =d _0 e^{f_d} =d _0 e^{d _5r}.
\end{equation}

In fact, galaxy structures depend only on the geometric curves, not
the choice of coordinate parameters.  Therefore, the following is
the general expression for equiangular spirals:
\begin{equation}
\left\{
\begin{array}{l}
x = e^{d_1 f(\lambda ) +d_2 g(\mu ) }\cos(d_3 f(\lambda ) +d_4 g(\mu ) ) ,   \\
y = e^{d_1 f(\lambda )+d_2 g(\mu ) }\sin(d_3 f(\lambda ) +d_4 g(\mu
) )
\end{array}
\right.
\end{equation}
where $f(\lambda ) ,g(\mu ) $ are arbitrary functions. All these
expressions give the same equiangular spirals and generate the same
exponential disks, independent of the choice of coordinate
parameters. That is, our rational structure which depends only on
the geometric proportion curves, is independent of the coordinate
system which describes the curves (coordinate invariance).

\begin{figure}
 \mbox{} \vspace{12.0cm} \includegraphics{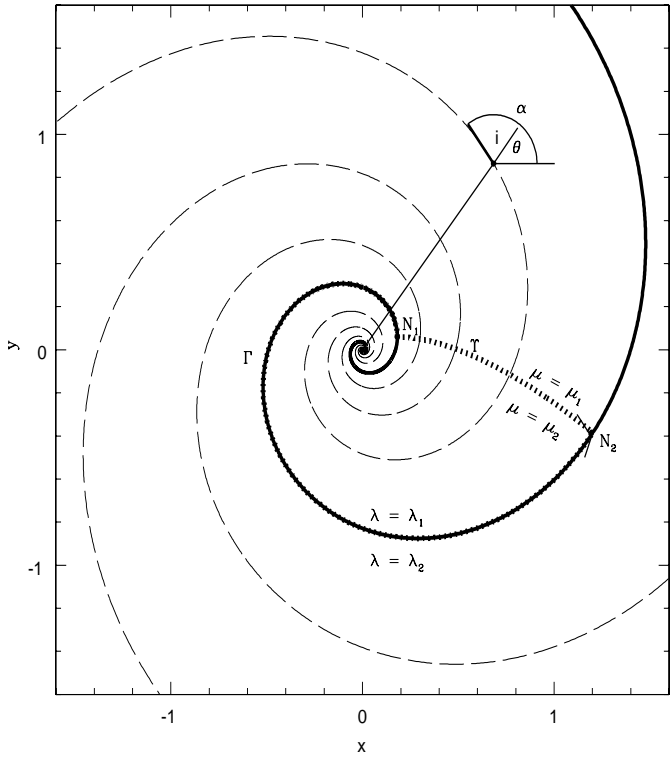}
\caption[]{
The orthogonal net of logarithmic spirals (14). The
angle $i$, at each position on the spiral, between its bending
direction and the disk radial direction is constant along the curve.
The closed curve  which consists of two sections of equiangular
spirals (thick dotted lines), demonstrates the closure condition
(see the Appendix A for details).
   }
\end{figure}

\newpage

\subsection{ The Second Example: Dual Handle Structure }

Now we study  galactic  bar model. A bar pattern is composed of two
or three dual handle structures which are generally aligned with
each other (spiral galaxy NGC 1365 is not the case). The dual handle
structure is determined by the following orthogonal curves of
confocal ellipses and hyperbolas:
\begin{equation}
\left\{
\begin{array}{l}
x=e^{\sigma } \cos \tau , \\
y=\sqrt{e^{2\sigma }+b_1^2 } \sin \tau ,  \\
-\infty < \sigma < +\infty,\; 0\le \tau <2\pi .
\end{array}
\right.
\end{equation}
where $b_1(>0)$ is a constant. The orthogonal coordinate system no
longer shares the coordinate lines with the polar coordinate system.
The coordinate lines are confocal ellipses and hyperbolas
(Fig.\,2). The distance between the two foci is $2b_1$ which
measures the distance between the two handles. The eccentric anomaly
of the ellipses is $\tau $.

Now we look for the logarithmic density ($f(\sigma , \tau )$) of the
dual handle structure determined by the coordinate system (27).
The corresponding arc-length derivatives  of the coordinate system
are (see (8) and (10)),
\begin{equation}
\begin{array}{l}
P=e^\sigma \sqrt {e^{2\sigma }+b^2_1\cos^2\tau }/\sqrt {e^{2\sigma }+b^2_1},       \\
Q = \sqrt {e^{2\sigma }+b^2_1\cos^2\tau } .
\end{array}
\end{equation}
The rational structure equation
\begin{equation}
u_b(\sigma ) P_\tau ^\prime - v_b(\tau )Q_\sigma ^\prime =0
\end{equation}
determines the corresponding directional derivatives of the
logarithmic density,
\begin{equation}
u_b (\sigma ) =b_2 e^\sigma \sqrt {e^{2\sigma }+b^2_1 }, \; v_b
(\tau ) = -b_2 b_1^2\sin\tau \cos\tau  .
\end{equation}
To get the logarithmic density we need to perform path integrations
of the similar formulas to (22). The result is
\begin{equation}
f_b (\sigma, \tau)= (b_2/3)( e^{2\sigma }+b_1^2\cos^2\tau)^{3/2}.  \\
\end{equation}
The inverse coordinate transformation of the formulas (27) is
easily found,
\begin{equation}
\begin{array}{l}
p(x,y)=e^{\sigma } =\sqrt {(r^2-b_1^2+\sqrt {(r^2-b_1^2)^2+4b_1^2x^2})/2},  \\
\cos\tau =xe^{-\sigma }=x/p(x,y)
\end{array}
\end{equation}
where $r^2=x^2+y^2$. Finally we find the density of the dual handle
structure,
\begin{equation}
\begin{array}{l}
f_b (x,y)= (b_2/3)( p^2(x,y) +b_1^2x^2/p^2(x,y))^{3/2},  \\
\rho _b =b_0 \exp(f_b (x,y))
\end{array}
\end{equation}
where $b _0$ is the dual handle density at the galaxy center. We
need to choose $b_2 <0 $ so that $f_b <0 $ and  $\rho _b \rightarrow
0$ when $r \rightarrow +\infty $. We can see that $b_0$ corresponds
to the central dual handle strength and $b_1$ corresponds to the
dual handle length while $b_2$ measures the density slope off the
dual handle. If we display the dual handle structure as a curved
surface in 3-dimensional space then we can see that the surface is
camelback-like shapes with two humps (i.e., handles).

Note that polar angle $\theta $ is not defined at the center $r=0$
for the polar coordinates. Similarly the eccentric anomaly $\tau $
is not defined for the coordinates (27) along the dual handle
central line of $2b_1$ length which is the coordinate line $\sigma
=-\infty $.

In fact, galaxy structures depend only on the geometric curves, not
the choice of coordinate parameters.  Therefore, the following is
the general expression for dual handle structures:
\begin{equation}
\left\{
\begin{array}{l}
x=e^{ f(\sigma ) } \cos g(\tau ), \\
y=\sqrt{e^{2 f(\sigma ) }+b_1^2 } \sin g(\tau )
\end{array}
\right.
\end{equation}
where $f(\lambda ) ,g(\mu ) $ are arbitrary functions. All these
expressions give the same dual handle structures, independent of the
choice of coordinate parameters. That is, our rational structure
which depends only on the geometric proportion curves, is
independent of the coordinate system which describes the curves
(coordinate invariance).

\begin{figure}
 \mbox{} \vspace{12.0cm} \includegraphics{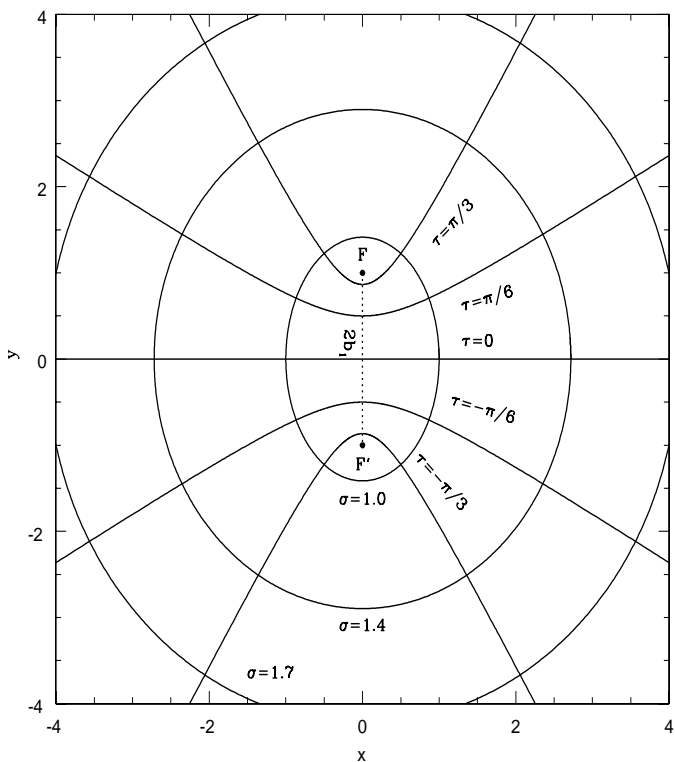}
\caption[]{
The orthogonal curves of confocal ellipses and
hyperbolas. The distance between the two foci, $F$ and $F^\prime $,
is $2b_1$ which measures the distance between the two handles.
   }
\end{figure}

\newpage

\section{ The Origin of Spiral Galaxies  }

\subsection{ Proposition\,1: Rational Structures are at Most Bilaterally Symmetric }

This is a mathematical proposition: any net of orthogonal curves is
either circularly symmetric with respect to the center point, or
bilaterally symmetric. I spent three years from 2002 to 2005 to look
for a net of orthogonal curves whose shape has odd symmetry. That
is, I wanted to find a rational structure which resembled a two-arm
spiral pattern like the spiral galaxy M51. The three-year study
indicates that a net of orthogonal curves is generally connected to
some complex analytical function. As you might know, a complex
analytical function always has two parts (real and imaginary) like
the formula (5), and leads to a net of orthogonal curves. However,
Complex analytical functions are very special ones which satisfy
some strong conditions like Cauchy integral theorem and formula.
From my experience, I do not find any complex analytical function
whose graph of the real or imaginary part has odd symmetry.
Therefore, I have the proposition that rational structures are at
most bilaterally symmetric. This is left for verification by able people.

Surprisingly, galaxy structures happen in the same way as indicated
in the following.

\subsection{ Coincidence\,1: Galaxy Patterns (except Arms) are at Most Bilaterally Symmetric }

Amazingly, any component of any galaxy pattern (except the arm
pattern) is either circularly symmetric, or bilaterally symmetric.
And my academic papers (He, 2005a, 2005b, and 2008) show that, except the arm structure,
any component of any galaxy structure (such as exponential disks,
galactic bars, even the whole elliptical galaxy) is a rational
structure. That is, any galaxy image (ignoring the arm) can be
fitted  identically to rational structures.

The following will present more and more cases of coincidences.
Therefore, they are not coincidence at all. They are the cosmic
truth, and my discovery is absolutely important.

\subsection{ Coincidence\,2: Dust and Irrationality }

It is the observational fact that spiral galaxies are full of dust
while elliptical galaxies have no dust. Why does this happen?
However, people have not found the appropriate answer since galaxies were
discovered more than 80 years ago.

Arm structure is neither circularly symmetric with respect to the
center point, nor bilaterally symmetric. Therefore, arms are not
rational structures. Arm pattern tends to be odd symmetry with
respect to the center. But we can never find such ``grand design''
arm pattern which presents the perfectly odd symmetry. In fact,
there are very different types of spiral structures. Some galaxies
are what we call ``Grand Design'' spirals,
meaning that they have a clearly outlined and well organized spiral
structure. Other galaxies, like NGC 4414 are called ``flocculent''
and have much harder to trace arms. Arms may not be called a
structure at all as compared with the exponential
disks, galactic bars, and even the whole elliptical galaxies which
demonstrate smooth and rational structures, .

Arm patterns exist only in spiral galaxies and they are weak
compared with the main disk structure of spiral galaxies. Therefore,
the presence of arm structure is the disturbance to the rational
structure. Because arm patterns exist only in spiral galaxies and
only spiral galaxies present dust, I comes to the critical
answer to the above question: any disturbance to rational structure
produces cosmic dust.

\subsection{ The Origin of Cosmic Dust: Impulsive Gravity }

From now on we call the disturbance to rational structure the cosmic
impulse. Therefore, the universe is originated not only rationally
but also impulsively. In the case of large-scale structures (that
is, galaxies), the impulse is demonstrated to be the disturbing
waves, i.e., the arm patterns.

We have seen that the rational structure is the main structure while
the disturbance to the main structure is always weak. That is,
rational force is the main one while the impulse is weak. In the
case of large-scale structures (that is, galaxies), the disturbing
waves try to achieve the minimal disturbance and, as a result, they
follow the proportion rows or columns of the rational structures.
The impulsive disturbance reveals the rational design of the
universe: proportion.

Because we live inside a galaxy (Milky Way galaxy) and we find no
other force  exists except gravity at our neighborhood, we conclude
that Newton or Einstein formulation of gravity is a partial result
of the universal ``gravity'' that not only presents rational sense
but also impulsive one as indicated at the large-scale galaxy
system.

\subsection{  A Simpler Argument that Spiral Galaxy Disk is a Rational Structure }

     The logarithmic stellar density of exponential disk is the circularly symmetric distribution of
numerical values about the center point, which decreases linearly in
the radial directions. Therefore, the magnitude of the directional
derivatives along radial directions is a global constant. An
equiangular spiral is the curve which makes a constant angle to the
local radial directions. Therefore, the magnitude of the directional
derivatives along the tangent directions of an equiangular spiral is
constant along the spiral. Accordingly the magnitude of the
directional derivatives along the perpendicular directions to the
equiangular spiral is also constant along the spiral. That is, you
walk along an equiangular spiral and the ratio of the density on
your left side to the immediate density on your right side is
constant along the curve. However, this constant is different from
the constant in the radial direction. They differ by a factor of the
cosine of the angle. Finally we have proved that exponential disk is
a rational structure.

   All curves which are orthogonal to a specific set of equiangular spirals in the clockwise direction
    are themselves
equiangular but in the counter-clockwise direction. These two sets
of spirals in opposite directions consist of the orthogonal net of
curves, and the exponential disk is the rational structure with
respect to these proportion  curves.

\subsection{ Coincidence\,3: Exponential Disk is Correlated with Equiangular Spiral}

The exponential disk of any spiral galaxy is a rational structure
which is circularly symmetric about the galaxy center. It is amazing
that its proportion curves are the equiangular spirals which are
precisely the curves represented by normal spiral galaxy arms. This
is another coincidence.

 Now we introduce more and more coincidences.

\subsection{ Coincidence\,4: The Only  Rational Structure which is not Circularly
Symmetric is Dual Handle Structure }

I have given the definition of rational structure. However, given an
arbitrary net of orthogonal curves, we are not always possible to
arrange a distribution of stellar density on the net to form a
rational structure. In fact, there are only a few types of
orthogonal curves which correspond to rational structures. Rational
structures are usually circularly symmetric about the center points.
The only rational structure we can find which is not circularly
symmetric, is the bilaterally symmetric structure, namely,
dual-handle structure.

We have two types of rational structures: exponential disks and
dual-handle structures. Adding the two structures together leads to
the barred pattern as we expected. It is amazing that only two kinds
of spiral galaxies are observed in the universe. One kind of spirals
are the normal spiral galaxies while the other kind are the barred
spiral ones. What is more surprising is that some barred galaxies do
show a set of symmetric enhancements at the ends of the stellar bar,
called the ansae or the ``handles'' of the bar (see upper-left panel
of Fig.\,3). This indicates that a bar itself is nothing but a set
of several pairs of ansae (handles). That is, bars are the
superposition of several aligned or misaligned dual-handle
structures. If the outer dual-handle structure is far more away from
the galaxy center then it demonstrates the pattern of ansae or
``handles'' of the bar.

\begin{figure}
 \mbox{} \vspace{12.0cm} \includegraphics{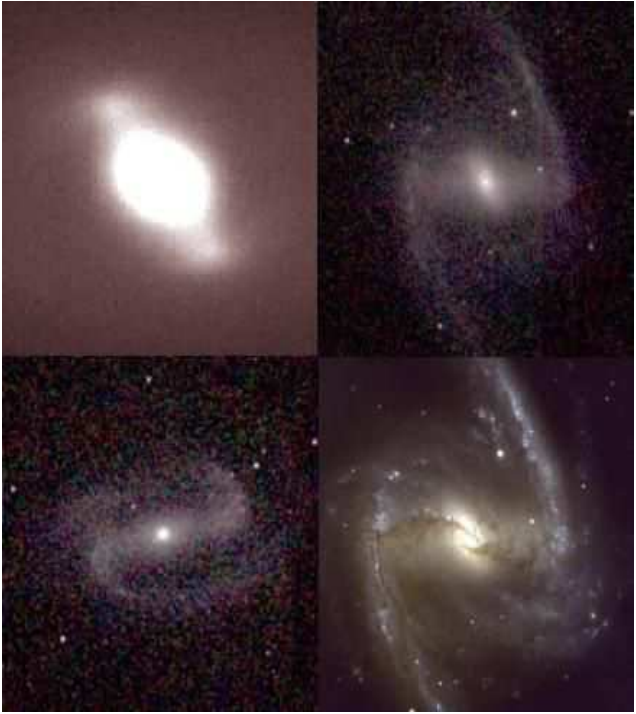}
\caption[]{
  Upper-left panel is galaxy NGC 2983 (image credit: Martinez-Valpuesta, Knapen, \&  Buta (2007)).
  Upper-right is the infrared
image of galaxy NGC 1365.
Lower-left is the infrared image of NGC 1300.
      Lower-right  is NGC 1365 (ultraviolet image, credit: European Southern Observatory).  }
\end{figure}

\newpage

\subsection{  Coincidence\,5: There are Barred Spiral Galaxies which
Present Two Nonparallel Bars }

The main structure of spiral galaxies is the exponential disk. When
the dual-handle structure (i.e., sub-bar) is near the galaxy center,
the superposition of the dual-handles to the bright disk center
presents a bar shape. This precisely explains the origin of galaxy
bars. A galaxy bar is usually composed of two or more sets of
aligned or misaligned dual-handle structures.  Surprisingly, there
are barred spiral galaxies which present two nonparallel bars (see
upper-right panel of Fig.\,3).

\subsection{  Coincidence\,6: Bar Structure is So Weak in the Outer Areas of Spiral
Galaxies that it is Ignored }

Compared with the exponential disk, the bar is observationally weak
structure. That is, bar structure is so weak in the outer areas of
spiral galaxies that it is ignored.  It is very surprising that the
theoretical calculation of dual-handle structure shows that it is
weak when compared with the disk (see Fig.\,4, 5, and 6).

We know that the disk density of spiral galaxies decreases outwards
exponentially, which is the numerical result obtained over 90 years
since the discovery of galaxies in the universe. Spiral galaxy disks
are thus called exponential disks. We add the dual-handle structure
to the  exponential disk for them to be the model of barred spiral
galaxies. If the density of dual-handle structure were comparable to
or stronger than the exponential disk in the far distances from the
galaxy center then our model would fail. That would suggest that the
main structure of spiral galaxies were not the exponential disk, a
result inconsistent with astronomical observation. The mathematical
result is that the density distribution of dual-handle structure
decreases outwards cubic-exponentially as shown below.

For large $r$, the first formula in (32) reduces to $r$, and the first formula in (33) approaches $(b_2/3)r^3$. Because $b_2 < 0$, we come to the conclusion that the density distribution of dual-handle structure decreases outwards cubic-exponentially.
In the distances far from
the galaxy center, the dual-handle structure is negligible when
compared with the disk. Mathematical result is consistent to
observation. This piece of coincidence alone proves that galaxies
are originated from proportion force.

\begin{figure}
 \mbox{} \vspace{12.0cm} \includegraphics{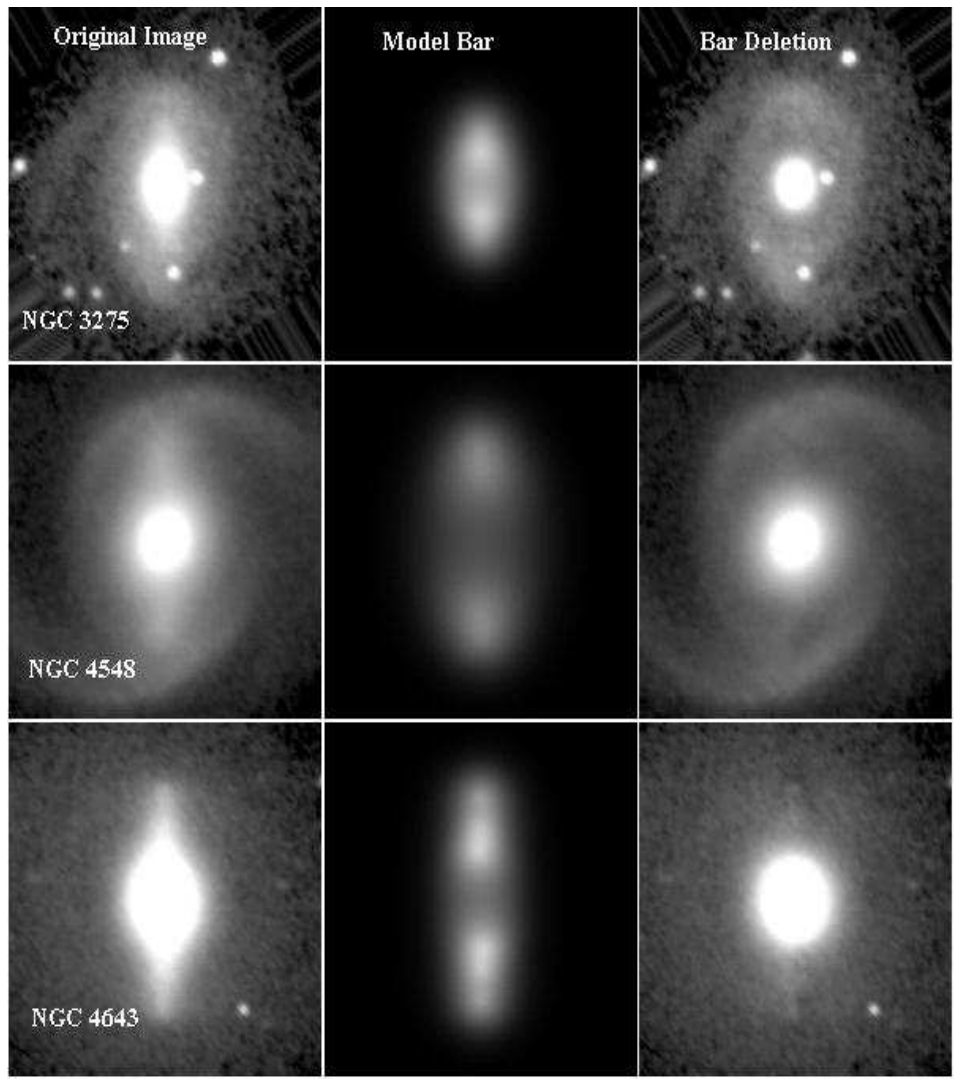}
\caption[]{
 The OSUBGS H-band images NGC 3275, 4548, 4643 (image
credit: Eskridge, {\it et al.} (2002)) minus our model bars
 respectively result in the disk and bulge images (bar deletion).}
\end{figure}

\subsection{ Coincidence\,7: The Arms of Barred Galaxies Spin around the Bar and
are No Longer the Equiangular Spirals }

With simple mathematical calculation we know that spiral-shaped
proportion curves exist in dual-handle structure. However, they are
not equiangular because they surround the central line of the
dual-handles (recall that the spirals in exponential disks are
equiangular and surround the center point). Two  proportion curves
which are oddly symmetric about the center point in dual-handle
structure make approximately elliptical shape and its long axis must
be parallel to the central line of the dual-handles. Surprisingly,
astronomical observations show that arms of barred spiral galaxies
do surround the middle lines of their bars, and they are not
equiangular, and the two arms make approximately elliptical shapes
with the long axes being parallel to the bar middle lines (see
Fig.\,3).

 \begin{figure}
 \mbox{} \vspace{12.0cm} \includegraphics{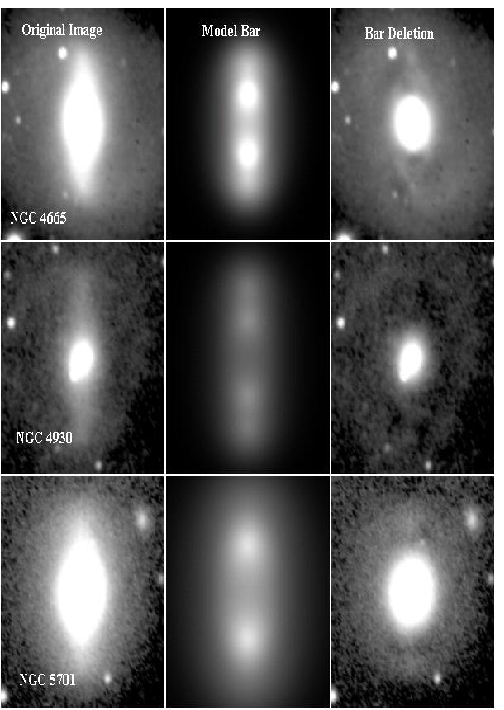}
\caption[]{
 The OSUBGS H-band images NGC 4665, 4930, 5701 (image
credit: Eskridge, {\it et al.} (2002)) minus our model bars
 respectively result in the disk and bulge images (bar deletion).}
\end{figure}

We proceed with more mathematical details. The proportion curves of
a dual-handle structure are all confocal ellipses and hyperbolas.
The two foci are the centers of the two handles. The distance
between the two foci is the length of the dual-handle structure.

If the length of the dual-handle structure is zero then the two foci
overlap to be the center of concentric circles and the above-said
proportion curves become the curves of polar coordinates. This
returns to the case of normal spiral galaxy disk. In fact, the polar
curves are also the proportion curves of normal spiral galaxies. In
other words, all polar curves are the limiting curves of equiangular
spirals.

 Back to the above-mentioned dual-handle structures.
Similarly, they have also open proportion spirals which make acute
angles to the above-mentioned ellipses and hyperbolas, and spin
around the central line of the dual-handle structure. However, the
spirals are no longer equiangular.
\subsection{  Coincidence\,8: Circular and Elliptical Rings }

We have already known that exponential disks have circular
proportion curves (one family of polar curves). Observationally,
some normal spiral galaxies do have closed arms which are circular,
called rings. Dual-handle structures also have closed proportion
curves which are ellipses whose long axes must be parallel to the
central lines of the dual-handles. Observationally, some barred
spiral galaxies do have closed rings which are ellipses and the long
axes are parallel to the galaxy bars.

\subsection{  Coincidence\,9:  Fitting Bar Images with Dual-handle Structures  }
          In fact, we can directly use dual-handle structures to fit into the bar images of
spiral galaxies. From the barred spiral galaxy images we subtract
the superposition of two or three sets of dual-handle structures and
see if
 the resulting images are exponential disks plus the weak arm structures.
If it works then we have further proved that barred spiral galaxies
are the superposition of exponential disks with dual-handle
structures. The first column of the Figures 4, 5 and 6 present
real galaxy images and the second column present the fitting bars.
The third column presents the result of the subtraction of the
second column from the first one. The results are very good: the
resulting images are exponential disks plus the weak arm structures.
In fact,
  I have made a piece of computer software which can be used to improve
the results.

\subsection{ Coincidence\,10:  Elliptical Galaxies are Completely
Rational Structures}

I have proved that elliptical galaxies are completely
rational structures in three-dimensions (He, 2005a and 2008). The proportion surfaces of
elliptical galaxies are the intersecting nets of orthogonal spheres,
where disturbing waves are difficult to form and spread. On the
other hand, spiral galaxies are two-dimensional and their proportion
curves are open spirals where disturbance waves are easy to form and
spread. Astronomical observations do show that arms do not exist in
elliptical galaxies.

The disturbance to rational structure leads to the formation of gas
and dust. New families of stars and planets are born to these gas
and dust. The star-planet families are short-lived. This happens
only in spiral galaxies.

\subsection{  It Seems that Fitting Galaxy Images may Tell the Physical
Distances of the Corresponding Galaxies  }
       Galaxies are very far away from the earth and human is really minute.
Therefore, we can not measure directly how far away the galaxies are
from Earth. It seems that fitting barred galaxy images with
dual-handle structures may tell the physical distances of the
corresponding galaxies. However, the result needs further
confirmation with the help of supercomputers.

\subsection{  Coincidence\,11:  Rational Galaxy Structures are Determined by Elementary Functions. }
It is amazing that rational galaxy structures are determined by the most elementary functions. Spiral galaxy disks and bars are determined by the complex exponential function $g= \exp w $ where $g = x+iy$ and $w= \lambda + i \mu$ (see the formula (14)). Elliptical galaxy structure is determined by the complex reciprocal function $g = 1/ w$ (see He (2008)). In fact, there are a few elementary functions which correspond to rational structures, and the nature chooses the most elementary functions to be its design of the universe.

\subsection{    Future Work and Coincidence\,12: Galaxy Nuclear Rings. }
There are lots of future theoretical and applied works to be done. One of them is the search of the condition under which the summation of two rational structures is still rational. Let $\rho _1 $ and $\rho _2 $ be the two rational structures. Their logarithmic densities are  $f_1 $ and $f _2 $ respectively. The logarithm of the summation is  $f $. It is straightforward to show that
$$
\nabla f = (\rho _1  \nabla f_1  + \rho _2  \nabla f_2 )/(\rho _1  + \rho _2 ).  \\
$$
That is, the gradient of summation is the summation of weighted gradients. Because galaxy disks take their maximum values at around the galaxy center and dual handle structures take the values away from the center, our above conclusions are still true. Furthermore, the proportion curves of the summation structure explain why there are nuclear rings or arms for some barred galaxies. An example is the galaxy NGC\,4314.

 \begin{figure}
 \mbox{} \vspace{12.0cm} \includegraphics{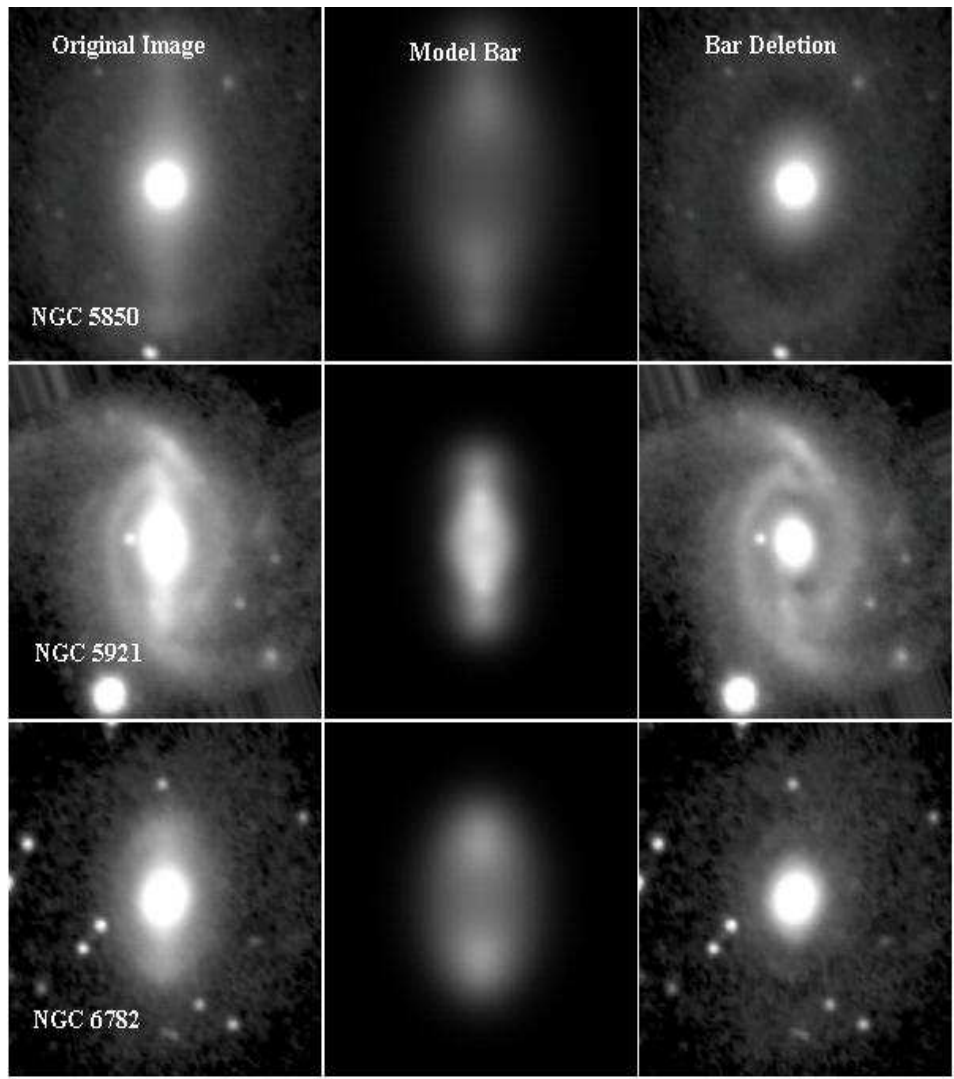}
\caption[]{
The OSUBGS H-band images  NGC 5850, 5921, 6782 (image credit:
 Eskridge, {\it et al.} (2002)) minus our model bars respectively result in the disk
and bulge images (bar deletion).
    }
\end{figure}

\newpage

\section{ Conclusion  }

1. The well-known fact can not be ignored that gravity is very very weak.
For example, it is $10^{-40}$ times weaker than the electricity between protons.
Therefore, humans in the foreseeing future can not design
a  physical precision experiment which can resolve the $10^{-40}$ strength of the earth's
gravitational field. That means we have not had a full understanding of gravity.
But scientists assume they had it and applied the preliminary results of Newton and Einstein
to the whole universe. This resembles the situation of
cycles and epicycles in the old geocentric model.

2. It is a fact that the results of Newton and Einstein deal with the motion of two-bodies.
When
applied to the free motion of many-bodies, the theories give chaotic results. However,
the universe has orderly motion. Whenever a problem involves free many bodies,
Newton and Einstein theories have no power. For example, the Bode law of planetary
distribution in the solar system has not been explained.

3. Newton and Einstein theories have no power for the explanation of natural structures.
Galaxy structure is the
simplest one in the observational world. Every one with common sense must
suggest that there exists a law on galaxy structure. Newton and Einstein theories
can not provide such law because they are the theories of two-bodies.
The law is very possibly the rationality explained in my article.

4. The mainstream model of the universe (the Big Bang theory) which is based on Newton and
Einstein theories, is being declined. A new article (Nieuwenhuizen, Gibson \& Schild (2009)) describes:
Nearly every month
new observations arise that pose further challenges to the
$\Lambda $CDM paradigm: Correlations in galaxy structures (Disney M. J. {\it et al.} (2008));
absence of baryon acoustic oscillations in galaxy-galaxy
correlations (Sylos Labini, Vasilyev \& Baryshev Yu (2009)); galaxies formed already when the universe
was 4 to 5 billion years old (Bouwens \& Illingworth (2006)); dwarf satellites that swarm
our own galaxy just like its stars (Metz M. {\it et al.} (2009)).
Observational data (Kroupa {\it et al.} (2010), Peebles \& Nusser (2010)) strongly suggest a paradigm shift for cosmology.

\appendix

 \section{ The Equiangular Coordinate System as a Uniquely Defined Mapping between Coordinate Spaces. }
We did not specify the variance domain $S$ on $(\lambda , \mu )$
coordinate plane on which the equiangular coordinate system (14)
is defined and maps it onto the whole $(x,y)$ coordinate plane. To
find the domain, we define two constants which are called periods,
\begin{equation}
\begin{array}{l}
\Delta _\lambda =\frac {2\pi d _2}{d _1d _4-d _3d _2} \;(>0),  \\
\Delta _\mu =\frac {2\pi d _1}{d _1d _4-d _3d _2} \; (>0).
\end{array}
\end{equation}
It is straightforward to show that the two periods satisfy the
following equations,
\begin{equation}
\begin{array}{l}
d_1\Delta _\lambda - d _2\Delta _\mu =0 , \\
-d _3 \Delta _\lambda  +d _4\Delta _\mu =2\pi.
\end{array}
\end{equation}

There are many such domains and now we want to prove the  statement
of coordinate periodicity that any vertical infinite band
$S_{\lambda _1} $ on $(\lambda , \mu )$ plane can be mapped onto the
whole $(x,y)$ plane of a galaxy disk,
\begin{equation}
S_{\lambda _1} :  \lambda _1 < \lambda < \lambda _1 +\Delta _\lambda
(=\lambda _2), \; -\infty < \mu < +\infty .
\end{equation}
where $\lambda _1 $  is arbitrary constant and the length of the
interval $(\lambda _1 , \lambda _2 )$ is the period,
\begin{equation}
 \lambda _2 = \lambda _1 +\Delta _\lambda .
\end{equation}
As indicated in Fig.\,1, it is equivalent to show that the two
different vertical boundary lines, $\lambda =\lambda _1$ and
$\lambda =\lambda _2$ on $(\lambda , \mu)$ plane are mapped to a
single curvilinear coordinate line on $(x, y)$ plane (see
Fig.\,1). This can be shown by three steps. We choose a closed
curve (thick dotted line in Fig.\,1) which consists of two
sections, one from the coordinate lines of first set ($\mu = $
constant $=\mu _1, \lambda _1 < \lambda < \lambda _2 $)  and the
other from the second set ($\lambda = $ constant $ =\lambda _1, \mu
_1 < \mu < \mu _2 $). The closed curve is called snail-shaped curve
(see Fig.\,1).

For the first step, we show that the arc-length derivative
$Q(\lambda, \mu )$ is uniquely defined along the snail-shaped curve.
Starting at the point $N_1 $ where the value of $Q$ is $Q(\lambda _1
, \mu _1)$, we have two ways to go to the point $N_2 $, one being
the section $\Gamma $ and the other $\Upsilon $. The corresponding
two values of $Q$ at $N_2 $ are $Q(\lambda _1 , \mu _2)$ and
$Q(\lambda _2 , \mu _1)$ respectively. The uniqueness of $Q$ means
that $Q(\lambda _1 , \mu _2) =Q(\lambda _2 , \mu _1)$. This is
guaranteed by the first equation in (36).

For the second step, the values of the tangent angle $\alpha
(\lambda , \mu)$ of the coordinate line $\lambda =\lambda _1$ (see
Fig.\,1) should be uniquely defined along the line within a
difference of $2\pi $. Starting at the point $N_1 $ where the value
of $\alpha $ is $\alpha (\lambda _1 , \mu _1)$, we have two ways to
go to the point $N_2 $, one being the section $\Gamma $ and the
other $\Upsilon $. The corresponding two values of $\alpha $ at $N_2
$ are $\alpha (\lambda _1 , \mu _2)$ and $\alpha (\lambda _2 , \mu
_1)$ respectively. Therefore,
\begin{equation}
 (\alpha (\lambda _1 , \mu _2)- \alpha (\lambda _1 , \mu _1))-
 (\alpha (\lambda _2 , \mu _1)- \alpha (\lambda _1 , \mu _1)) =2\pi.
\end{equation}
Now we need calculate the value of $\alpha (\lambda , \mu)$. Its
calculation is straightforward with formulas (14),
\begin{equation}
\tan \alpha = \frac {y^\prime _\mu }{x^\prime _\mu }= \frac
{d_2\sin\theta +d_4\cos\theta }{ d_2\cos\theta -d_4\sin\theta }.
\end{equation}
Defining
\begin{equation}
\cos \theta _0 =\frac {d_2 }{ \sqrt{d_2^2+d_4^2} },\; \sin \theta _0
=\frac {d_4 }{ \sqrt{d_2^2+d_4^2} },
\end{equation}
then we have
\begin{equation}
\tan \alpha = \tan (\theta +\theta _0 ).
\end{equation}
Finally we find the value of the tangent angle of the coordinate
line $\lambda =$ constant
\begin{equation}
\alpha =\theta +\theta _0 = d_3\lambda +d_4\mu +\theta _0.
\end{equation}
Substituting the value into the equation (39), we find that the
equation reduces to  the second equation in (36) which is already
proved. Note that, in Fig.\,1, $\alpha $ is the tangent angle of
the coordinate line $\lambda =$ constant at a point $(x,y)$ and
$\theta $ is the polar angle of the same point. Their difference is
the pitch angle of the coordinate line $\lambda =$ constant.
Therefore, we find the pitch angle $i$ to be,
\begin{equation}
i=\alpha -\theta =\theta _0
\end{equation}
which is constant over the whole plane. We have proved that our
coordinate lines (proportion curves) are equiangular spirals.

For the third step, we need prove that the snail-shaped curve is
really closed. To see the case, we change the coordinate
transformation equations (14) into different form, making use of
the formulas (17) and (43),
\begin{equation}
\begin{array}{l}
x = Q(\lambda ,\mu)\cos(\alpha -\theta _0 ) /\sqrt{d_2^2 +d_4^2 },    \\
y = Q(\lambda ,\mu)\sin(\alpha -\theta _0 ) /\sqrt{d_2^2 +d_4^2 } .
\end{array}
\end{equation}
Because $Q(\lambda ,\mu)$ and $\alpha $ are uniquely defined and
both $\theta _0 $ and $\surd (d_2^2 +d_4^2 )$ are constants, $x,y$
are uniquely defined. That is, the snail-shaped curve is really
closed.

Therefore, we have shown that the two different vertical boundary
lines, $\lambda =\lambda _1$ and $\lambda =\lambda _2$, on $(\lambda
, \mu)$ plane are mapped  to a single curvilinear coordinate line on
$(x, y)$ plane (see Fig.\,1). This is equivalent to say that the
vertical infinite band $S_{\lambda _1} $  on $(\lambda , \mu )$
plane (see (36)) is mapped onto the whole $(x,y)$ plane of the
galaxy disk. Because $\lambda _1$ is arbitrarily chosen, we have
shown that the coordinate transformation (14) is periodic in that
each interval of $\lambda $ of length (period) $\Delta _\lambda $
with $-\infty < \mu < +\infty$ is mapped onto whole $(x,y)$ plane.
The periodicity of $\mu $ can be similarly proved.

\end{document}